\documentclass[sigconf]{acmart}
\AtBeginDocument{%
  \providecommand\BibTeX{{%
    \normalfont B\kern-0.5em{\scshape i\kern-0.25em b}\kern-0.8em\TeX}}}

\copyrightyear{}
\acmYear{}
\acmDOI{}

\acmConference[Preprint]{}{}{}
\acmBooktitle{}
\acmPrice{}
\acmISBN{}

\newcommand{\hyphen}{\mathchar`-}
\newcommand{\mathfont}{\mathsf}
\newcommand{\Freq}{F}

\newcommand{\minicirc}{\triangleleft}
\newcommand{\relu}{\mathfont{ReLU}}
\newcommand{\highway}{\mathfont{H}}
\newcommand{\trans}{^\mathfont{T}}
\newcommand{\dilate}{\delta}
\newcommand{\channels}{d}
\newcommand{\cnnbasic}[4]{#4\mathfont{C}{}^{{#1}}_{{#2} \star {#3}}}
\newcommand{\resizetensor}[4]{\cnnbasic{#1}{#2}{#3}{#4}}
\newcommand{\cnn}[4]{\cnnbasic{#1}{#2}{#3}{\highway}}
\newcommand{\dcnn}[4]{#4\mathfont{D}{{}^{{#1}}_{{#2} \star {#3}} } }
\newcommand{\att}[3]{\mathfont{Att}({#1},{#2},{#3})}

\begin{document}

\title{Analysis and Assessment of Controllability of an Expressive Deep Learning-based TTS system}


\author{No\'e Tits}
\affiliation{%
  \institution{Numediart Institute - UMONS}
  \streetaddress{31 Bd Dolez}
  \city{Mons}
  \country{Belgium}}
\email{noe.tits@umons.ac.be}

\author{Kevin El Haddad}
\affiliation{%
  \institution{Numediart Institute - UMONS}
  \streetaddress{31 Bd Dolez}
  \city{Mons}
  \country{Belgium}}
\email{kevin.elhaddad@umons.ac.be}

\author{Thierry Dutoit}
\affiliation{%
  \institution{Numediart Institute - UMONS}
  \streetaddress{31 Bd Dolez}
  \city{Mons}
  \country{Belgium}}
\email{thierry.dutoit@umons.ac.be}

\renewcommand{\shortauthors}{Tits et al.}

\begin{abstract}
In this paper, we study the controllability of an Expressive TTS system trained on a dataset for a continuous control. The dataset is the Blizzard 2013 dataset based on audiobooks read by a female speaker containing a great variability in  styles and expressiveness.
Controllability is evaluated with both an objective and a subjective experiment.
The objective assessment is based on a measure of correlation between acoustic features and the dimensions of the latent space representing expressiveness.
The subjective assessment is based on a perceptual experiment in which users are shown an interface for Controllable Expressive TTS and asked to retrieve a synthetic utterance whose expressiveness subjectively corresponds to that a reference utterance.
\end{abstract}

\ccsdesc[500]{Human-centered computing~Walkthrough evaluations}

\keywords{Deep Learning; Speech Synthesis; Style Interpolation; Perception}


\maketitle

\section{Introduction}
%
%
%
%


Text-To-Speech (TTS) frameworks, that generate speech from textual information, have been around for a few decades and have improved lately with the coming of new AI methods, e.g., Deep Neural Networks (DNN). 
Commercial products provide user--friendly DNN-based speech synthesis systems. Such recent systems offer an excellent quality of speech obtained by analyzing tens of hours of neutral speech which often fail to convey any emotional contents.
The task looked by scientists today has evolved towards the field of expressive speech synthesis~\cite{emotional_speech_synthesis-14-Burkhardt}. 

The aim of this task is to create, not an average voice, but specific voices, with particular grain and extraordinary potential with regards to expressiveness. This will make it possible to make virtual agents behave in a characteristic way, and hence to improve the nature of the interaction with a machine, by getting closer to a human-human interaction. 
It remains to find good ways to control such expressiveness characteristics.

The paper is organized as follows: 
\begin{itemize}
    
\item  related work is presented in Section~\ref{related_work}; 
\item  Section~\ref{sec:system} describes the proposed system for controllable expressive speech synthesis;
Section~\ref{sec:post_analysis}  presents the methodology that allows  to  discover  the  trends of audio features in  the  latent space;
\item Section~\ref{sec:objective_experiment} presents objective results using this methodology, and results regarding the acoustic quality with measures of errors between generated acoustic features and ground truth;
\item the procedure and results of the perceptual experiment is described in
Section~\ref{sec:subjective_experiment}; 
\item finally we conclude and detail our plans for future work in Section~\ref{sec:ccl}.

To obtain the results of the experiments of this paper, the software presented in~\cite{ICE_TALK2-21-tits} was used. It is available online\footnote{\url{https://github.com/noetits/ICE-Talk}}
A code capsule\footnote{\url{https://doi.org/10.24433/CO.1645822.v1}} provides an example of use of the software with LJ-speech dataset~\cite{ljspeech-17-keithito} which in the public domain.

\end{itemize}

\section{Related work \& challenges} 


\label{related_work} 

The voice quality and the number of control parameters depend on the synthesis technique used~\cite{emotional_speech_synthesis-14-Burkhardt}. These parameters allow creating variations in the voice. The number of parameters is subsequently important for the generation of expressive speech. 

Historically, there have been different approaches to expressive speech synthesis.
Formant synthesis can control numerous parameters, however the generated voice is unnatural. Synthesizers using the concatenation of voice segments reach a higher naturalness, however this technique give few control possibilities.

The first statistical approaches using Hidden Markov Models (HMMs)~\cite{hmms_to_dnns-16-watts} allowed to achieve both a fair naturalness and a control of numerous parameters~\cite{statistical_param_speech_syn-09-zen}. The latest statistical approaches use DNN~\cite{stat_param_speech_synthesis_dnn-13-zen} and was the premise of new speech synthesis frameworks, for example, WaveNet~\cite{wavenet-16-vandenoord} and Tacotron~\cite{tacotron-17-wang}, referred to as Deep Learning-based TTS.

Regarding the controllable part of TTS framework, a significant issue is the labeling of speech information with style or emotion data. 
Late investigations have been directed into unsupervised strategies for how to accomplish expressive speech synthesis without the need for annotations. 

A task related to controllable expressive speech synthesis is the prosody transfer task for which the goal is to synthesize speech from text with a prosody similar to another audio reference. A common characteristic of both tasks is the need for a representation of expressiveness.
However, for controllable speech synthesis, this representation should be a good summary of expressiveness information, i.e., it should be interpretable. A low dimension would help the interpretability.
For prosody transfer, the representation should be as accurate and precise as possible.

In~\cite{tacotron_prosody-18-skerry}, the authors present a prosody transfer system extending the Tacotron speech synthesis architecture. This extension learns a latent embedding space by encoding audio into a vector that conditions Tacotron along with the text representation. These latent embeddings model the remaining variation in speech signals after accounting for variation due to phonetics, speaker identity, and channel effects.

In~\cite{fine_grained_prosody_transfer-19-drugman}, they propose a supervised approach that use a time-dependent prosody representation based on F0 and the first mel generalized ceptral coefficient (representing energy).
They use of a dedicated attention module and a VAE to be able to concatenate this information to linguistic encodings. This allows for a fine-grained prosody transfer instead of a sentence level prosody information.

CopyCat~\cite{copycat-20-drugman} addresses the problem of speaker leakage in many-to-many prosody transfer. This problem occurs when the voice of the reference sample can be heard in the resulting synthesized speech while it should only transfer prosody and not speaker identity.
They are able to reduce the phenomenon with a novel reference encoder architecture that captures temporal prosodic representations robust to speaker leakage.

Concerning controllable speech synthesis, \cite{expressive_speech_vae-18-akuzawa} proposed to use a VAE and deploy a speech synthesis system that combines VAE with VoiceLoop~\cite{voiceloop-17-taigman}.
Some other researches have used the concept of VAE~\cite{generative_controllable_speech-18-hsu,unsupervised_controllable_speech-18-henter} for controllable speech synthesis. In~\cite{generative_controllable_speech-18-hsu}, the authors combine VAE and GMM and call it GMVAE. 
For more details concerning the different variants of such methods, an in-depth study of methods for unsupervised learning of control in speech synthesis is given in~\cite{unsupervised_controllable_speech-18-henter}.  
These works show that it is possible to build a latent space leading to variables that can be used to control the style of synthesized speech.

In~\cite{tacotron_style_tokens-18-wang}, the authors show an example of spectrograms corresponding to a text synthesized with different rhythms, speaking rates and F0. However these works do not provide insights about the relationships between the computed latent spaces and the controllable audio characteristics. 

Different supervised approaches were also proposed to control specific characteristics of expressiveness~\cite{seq2seq_prosody_modif-19-shechtman, controllable_NTTS-20-raitio}. In these approaches, it is necessary to make a choice of control parameters, a priori, such as pitch, pitch range, phone duration, energy, and spectral tilt. This reduces the possibilities of the controllability of the speech synthesis system.

A shortcoming of these investigations is that they do not give insights about the extent to which the system is controllable from an objective and subjective point of view. We intend to fill this gap.



 

\section{System}

\subsection{DCTTS}

As our system relies on DCTTS~\cite{dctts-17-tachibana}, the details of the different blocks are given in Figure~\ref{fig:network-detail}. We use the notations introduced in~\cite{dctts-17-tachibana} in which the reader can find more details if needed:

\begin{itemize}
    \item 1D convolution:
    $\cnnbasic{o \gets i}{k}{\dilate}{} (X)$,
    where
    $ i $ is the sizes of input channel,
    $ o $ is the sizes of output channel,
    $ k $ is the size of kernel,
    $ \dilate $ is the dilation factor,
    and an argument $X$ is a tensor having three dimensions ({\it batch, channel, temporal}).
    The stride is $1$.

        \item 1D {\it deconvolution}:
    layer as $\dcnn{o \gets i}{k}{\dilate}{} (X)$.
    The stride is $2$.
    
        \item Layer composition operator: $\cdot \minicirc \cdot$, and
    
        \item Networks :
    $\mathfont{F} \minicirc \mathfont{ReLU} \minicirc \mathfont{G} (X)
        := \mathfont{F}(\mathfont{ReLU} ( \mathfont{G} (X))),$
    and
    $(\mathfont{F} \minicirc \mathfont{G})^2(X)
        := \mathfont{F} \minicirc \mathfont{G} \minicirc \mathfont{F} \minicirc \mathfont{G} (X)$, etc.
        
        \item $\mathfont{ReLU}$ is an element-wise activation function defined by $\mathfont{ReLU}(x) = \max(x, 0)$.
        
        \item Highway Network-like gated activation,
    $\mathfont{Highway}(X; \mathfont{L}) = \sigma(H_1) \odot H_2 + (1 - \sigma(H_1)) \odot X,
    $
    
    where $H_1, H_2$ are output by a layer $\mathfont{L}$ as $[H_1, H_2]= \mathfont{L}(X)$.
    The operator $\odot$ is the element-wise multiplication
    
    $
    \cnn{\channels \gets \channels}{k}{\dilate}{}(X) := \mathfont{Highway} (X; \cnnbasic{2d \gets d}{k}{\dilate}{})$.

\end{itemize}

\begin{figure}[t]
    %
    \begin{minipage}[t]{0.99\linewidth}
            \input{figures/networks}
    \end{minipage}
    \caption{{ Details of DCTTS architecture \cite{dctts-17-tachibana}}}\label{fig:network-detail}
\end{figure}

\label{sec:system}
\begin{figure*}[h]
\centering
\includegraphics[scale=0.2, width = 1\textwidth]{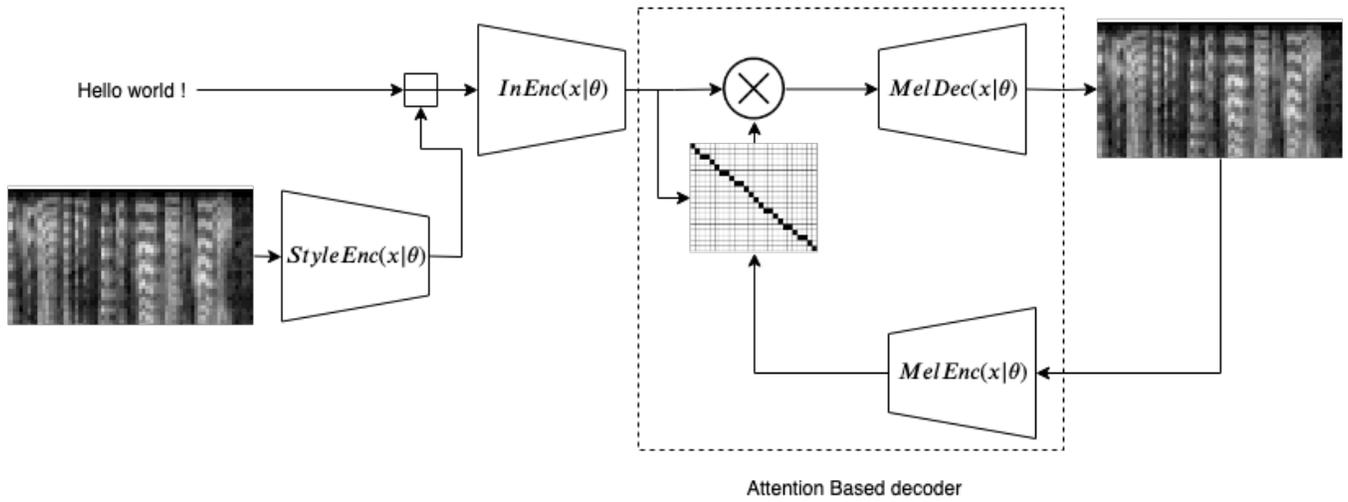}
\caption{Block diagram of the system}
\label{system}
\end{figure*}

\subsection{Controllable Expressive TTS}

The system is a Deep Learning-based TTS system that was modified to enable a control on acoustic features through a latent representation. The basis system Deep Convolutional TTS (DCTTS)~\cite{dctts-17-tachibana}. 


Figure~\ref{system} shows a diagram of the whole system. The basis DCTTS system is constituted of the $InEnc$, the Attention based decoder comprising $MelEnc$ and $MelDec$. 

For the latent space design, the $StyleEnc$ network was added. It consists of a stack of 1D convolutional layers similar to the $MelEnc$, followed by an average pooling. This operation enforces to encode time independent information. It can thus contain information about statistics of prosody such as pitch average, average speaking rate, but not a pitch evolution.
The latent vector at the output is the representation of expressiveness. This vector is then broadcast-concatenated.

This system was compared to other in~\cite{visualization-19-tits}. This comparison was done by training the system with a single speaker dataset with several speaking styles given by an actor and recorded in studio. 


In this paper we study the control of this system trained on dataset with which we hope will enable a continuous control of expressiveness. The dataset is the Blizzard2013 dataset~\footnote{\url{https://www.cstr.ed.ac.uk/projects/blizzard/2013/lessac_blizzard2013/}} by Catherine Byers based on audio book with a great variability in vocal expressions and therefore in acoustic features.

The latent space is designed to represent this acoustic variability and as a control to the output. It allows this without having any annotation regarding the expressiveness, emotion, style because the representation is learned during the training of the architecture.










\section{Post-Analysis for interpretation of Latent Spaces}
\label{sec:post_analysis}
In this section, we explain the method presented in~\cite{visualization-19-tits}. The methodology allows to discover the trends in the latent space. It can be done in the original Latent Space or in a reduced version of it.

The goal is to map mel-spectrograms into a space which is hopefully organized to represent the acoustic variability of the speech dataset.

To analyze the trends of acoustic features in latent spaces, we compute the direction of greatest variation in the space. For each feature of a set, we perform a linear regression using the point in the latent space and the feature computed from the corresponding file in the dataset.


\begin{figure}[!t]
\centering
\includegraphics[width=0.7\columnwidth]{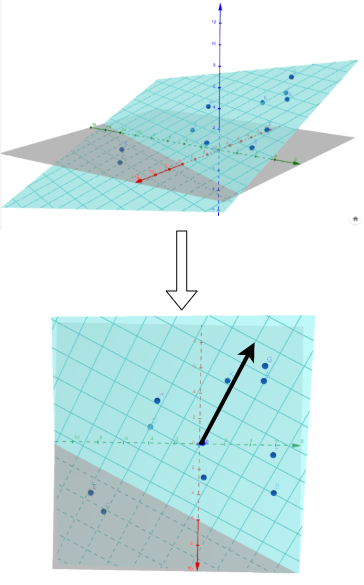}
\caption{Gradient of the hyperplane corresponding to the greatest slope}
\label{fig:gradients}
\end{figure}




The steps are the following:
\begin{itemize}
    \item The mel-spectrogram is encoded to vector of length 8 that contains expressiveness information. This vector is computed for each utterance of the dataset. 
    \item Dimensonality reduction is used to have a ensemble of 2D vectors instead. Figure~\ref{fig:gradients} shows a scatter plot of these 2D points. 
    \item Then a trend is extracted for each audio feature. For, e.g. $F0_{mean}$, its value is computed for each utterance of the dataset. We obtain therefore a $F0_{mean}$ corresponding to each 2D-points $(x,y)$ of the scatter plot. 
    \item We approximate the plane $$F0 = f(x,y) =ax+by+c$$
    \item To assess that this plane $f(x,y)$ is a good approximation of $F0_{mean}$, implying a linear relation between a direction of the space and $F0_{mean}$, we compute the correlation between the approximations $f(x,y)$ with the ground truth values of $F0_{mean}$.
    \item If we compute the gradient of the plane (which is in fact $(a,b)$ ), we have the direction of the greatest slope, that is plot in blue.
\end{itemize}



This representation is useful for a perspective of interface for controllable speech synthesis system on which are represented the trends of audio features in the space.

\section{Objective Experiments}
\label{sec:objective_experiment}

First we follow the methodology presented in the previous section to extract the directions in the latent space corresponding to acoustic features of eGeMAPS feature set and quantify to which extent they are related by computing an Absolute Pearson Correlation Coefficient (APCC).


This feature set is based on Low-level descriptors (F0, formants, mfcc, etc.) to which are applied statistics for the utterance (mean, normalized standard deviation, percentiles).
All  functionals  are  applied  to  voiced  regions  only (non-zero F0). For MFCCs, there is also a version applied to all regions (voiced and unvoiced).

These features are defined in~\cite{egemaps-16-eyben} as follows:
\begin{itemize}
    \item F0: logarithmic F0 on  a  semitone  frequency scale, starting at 27.5 Hz (semitone 0)
    \item F1-3: Formants 1 to 3 centre frequencies
    \item Alpha  Ratio:  ratio  of  the  summed  energy  from 50-1000 Hz and 1-5 kHz
    \item Hammarberg Index: ratio of the strongest energy peak in the 0-2 kHz region to the strongest peak in the 2-5 kHz region.
    \item Spectral Slope 0-500 Hz and 500-1500 Hz: linear regression  slope  of  the  logarithmic  power  spectrum within the two given bands.
    \item mfcc1-4: Mel-Frequency  Cepstral  Coefficients 1 to 4
\end{itemize}

To objectively measure the ability of the system to control voice characteristics, we do a sampling in the latent spaces and verify that the directions control what we want them to control.

Then we assess the quality of the synthesis using some objective measures.



\subsection{Quantitative Analysis}
\subsubsection{Correlation Analysis}
To visualize acoustic trends, it would be useful to have a small number of features that gives a good overview. To extract a subset of the list, we apply a feature selection with a filtering method based on Pearson's correlation coefficient.
The idea is to investigate correlations between audio features themselves to exclude redundant features and select a subset.

The steps are the following:
\begin{itemize}
    \item features are sorted by APCC in decreasing order;
    \item for each feature, APCC with previous features are computed;
    \item if the maximum of these $inter-features-APCCs >0.8$, the feature is eliminated;
    \item finally, only features that have a $prediction-APCC>0.3$ are kept.
\end{itemize}

These limits are arbitrary and can be changed to filter more or less features from the list.








In Table~\ref{best_APCC_blizzard}, we show the results of the APCC for Blizzard dataset and show the plot of gradients. It can be noted that F0 median is the most predictable feature from the latent space. The feature selection method highlight a set of 17 diverse features that have an APCC $>0.3$.

\begin{table}[ht]
\caption{APCC values between the best possible hyperplane of the 
latent space and audio features of the eGeMAPS feature set.}
\label{best_APCC_blizzard}
\begin{center}
\begin{tabular}{|c|c|c|c|c|c|c|c|c|}
\hline
{} &         APCC \\
\hline
F0 percentile50.0 &  0.723824 \\
mfcc1V mean                        &  0.619622 \\
mfcc1 mean                           &  0.554794 \\
logRelF0-H1-A3 mean                &  0.493066 \\
mfcc4V mean                        &  0.492359 \\
HNRdBACF mean                      &  0.482579 \\
F1amplitudeLogRelF0 mean           &  0.473154 \\
slopeV0-500 mean                   &  0.420381 \\
StddevVoicedSegmentLengthSec               &  0.388952 \\
F3amplitudeLogRelF0 stddevNorm      &  0.360528 \\
mfcc2V mean                        &  0.360144 \\
hammarbergIndexV mean              &  0.356113 \\
mfcc1V stddevNorm                   &  0.350918 \\
loudness meanFallingSlope             &  0.350369 \\
loudness percentile20.0               &  0.340973 \\
loudness meanRisingSlope              &  0.323489 \\
F1frequency mean                   &  0.318096 \\

\hline
\end{tabular}
\end{center}
\end{table}







\subsubsection{Distortion analysis: a comparison with typical seq2seq}

To compare the synthesis performance of the proposed method with a typical seq2seq method, we compare objective measures used in expressive speech synthesis.
These measures compute an error between acoustic features of a reference and a prediction of the model. There exist different types of objective measures that intend to quantify the distortion induced by a system on audio quality or prosody. In this work, we use the following objective measures:\\

\begin{itemize}
    \item MCD~\cite{MCD} measuring speech quality: \\ $\textrm{MCD}_K = \frac{1}{T}\sum_{t=0}^{T-1}\sqrt{\sum_{k=1}^K \left(c_{t,k} - c_{t,k}'\right)^2}$
    \item VDE~\cite{VDE}: $\textrm{VDE} = \frac{\sum_{t=0}^{T-1} 1[v_t \neq v_t']}{T}$
    \item F0 MSE measuring a distance between F0 contours of prediction and ground truth: $\textrm{F0\_MSE} = \frac{1}{T}\sum_{t=0}^{T-1}{ \left(F_{0t} - F_{0t}'\right)^2}$
    \item lF0 MSE, similar to previous one in logarithmic scale: $\textrm{lF0\_MSE} = \frac{1}{T}\sum_{t=0}^{T-1}{ \left(\log{F_{0t}} - \log{F_{0t}'}\right)^2}$
\end{itemize}

Some works use DTW to align acoustic features before computing a distance. The problem with this method is that it modifies the rythm and speed of the sentence.
However computing a distance on acoustic features that are shifted completely distorts the results, therefore, it is needed to apply a translation on acoustic features and take the smallest possible distance.
We thus report measures with DTW and with shift only in Table~\ref{distortion_typical_tts} for the original DCTTS and Table~\ref{distortion_unsup_tts} for the proposed Unsupervised version of DCTTS. \\

\begin{table}[ht]
\caption{Objective measures for the typical TTS system.}
\label{distortion_typical_tts}
\begin{tabular}{lrrrr}
\hline
{} &        MCD &       VDE &   lF0\_MSE &       F0\_MSE \\
\hline
DTW   &   9.973914 &  0.015488 &  0.436348 &  1219.128507 \\
shift &  13.331841 &  0.236024 &  6.283607 &  9481.103150 \\
\hline
\end{tabular}
\end{table}

\begin{table}[ht]
\caption{Objective measures for the proposed Unsupervised TTS system.}
\label{distortion_unsup_tts}
\begin{tabular}{lrrrr}
\hline
{} &        MCD &       VDE &   lF0\_MSE &       F0\_MSE \\
\hline
DTW   &   9.624296 &  0.009699 &  0.311803 &   957.290498 \\
shift &  12.675999 &  0.218258 &  5.838526 &  8931.599942 \\
\hline
\end{tabular}
\end{table}

\subsection{Qualitative Analysis}
\begin{figure}[h]
  \includegraphics[width=\linewidth]{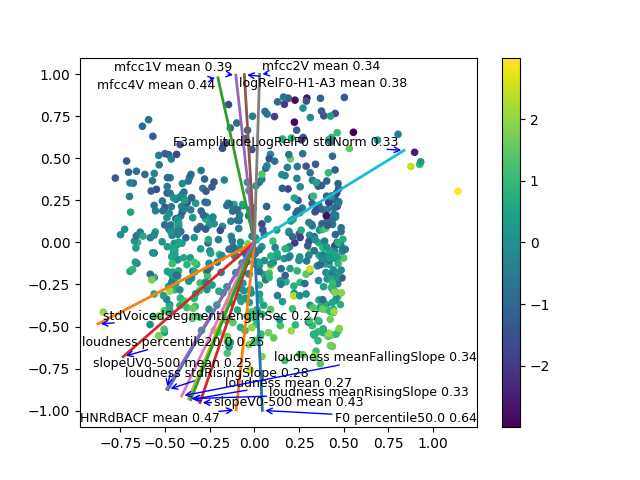}
   \includegraphics[width=\linewidth]{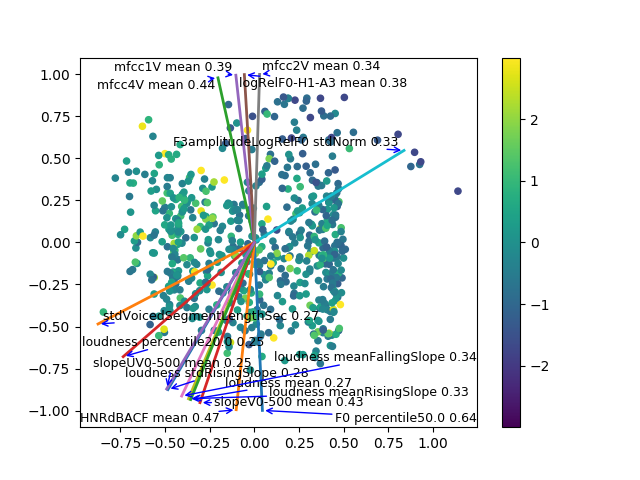}
  \caption{Reduced latent space with directions of gradients of features. 
  The color of each point is the value of F0 median (top) and Voiced segment lengths standard deviation (bottom).
  }
  \label{fig:gradients_continuous}
\end{figure}

In Figure~\ref{fig:gradients_continuous}, we show a scatter plot of the reduced latent space with the feature gradients. Each point corresponds to one utterance encoding and reduced to two dimensions. The color of these points is mapped to the values of an acoustic feature to be able to visualize how the gradients are linked to the evolution of the acoustic features. Two examples are shown for F0 median and standard deviation of voiced segment length, i.e., the duration of voiced sounds which is linked to the speaking rate. \\

We can observe that the direction of the gradients follows well the general trend of the corresponding acoustic feature. As the correlation values indicate, F0 median has an evolution closer to a linear evolution in the direction of the gradient rather than for voiced segment lengths standard deviation.

\section{Subjective Experiment}
\label{sec:subjective_experiment}

\subsection{Methodology}

An experiment was designed to assess the extent to which participants would be able to produce a desired expressiveness for a synthesized utterance, i.e., a methodology for evaluating the controllability of the expressiveness.

For this purpose, participants were asked to use the 2D interface to produce the same expressiveness as in a given reference. 
We assume that if participants are able to locate in the space the expressiveness corresponding to the reference, it means they are able to use this interface to find the expressiveness they have in mind.

The experiment contains two variants: in the first, the text of the reference and 2D space sentences are the same, while in the second, they are different. In the first one the participant can rely on the intonation and specific details of a sentence while in the second, he has to use a more abstract notion of expressiveness of a sentence.

The experiment is designed to avoid choosing a set of different characteristics or style categories, and letting the participant of the experiment judge how close the vocal characteristics of a synthesized sentence is to a reference.

The procedure for preparing the experiment is as follows:
\begin{itemize}
    \item The model trained with Blizzard2013 dataset is used to generate a latent space with continuous variations of expressiveness as presented in Section~\ref{sec:system}.
    \item In the 2D interface, we sample a set of points inside the region of the space in which the dataset points are located. 
    The limits of the rectangle are defined by projecting sentences of the whole dataset in the 2D space with PCA and selecting $x_{min}$, $x_{max}$, $y_{min}$, $y_{max}$ of all points. In other words, we use the smallest rectangle containing the dataset points.
    We use a resolution of 100 for $x$ and $y$ axes, making a total of 10000 points in the space.
    \item This set of 2D points is projected to the 8D latent space of the trained unsupervised model with inverse PCA. The 8D vectors will then be fed to the model for synthesis.
    \item 5 different texts are used to synthesize the experiment materials. This makes a total of 50000 expressive sentences synthesized with the model.
\end{itemize}

\vspace{1cm}

The listening test was implemented with the help of turkle\footnote{\url{https://github.com/hltcoe/turkle}}, which is an open-source web server equivalent to Amazon's Mechanical Turk that one can host on a server or run on a local computer.
We can ask questions with an HTML template that includes in this case an interface implemented in HTML/javascript.

During the perceptual experiment, a reference sentence coming from the 50000 sentences is provided to the participants.
We provide the interface allowing a participant to click in the latent space and choose what is the point that is in his opinion the closest to the reference in terms of expressiveness.


The instructions shown to participants are the following:
\begin{itemize}
    \item First, before the experiment, to illustrate what kind of task it will contain and familiarize you with it, here is a link to a demonstration interface:\\
    \url{https://jsfiddle.net/g9aos1dz/show}
    \item You can choose the sentence and you have a 2D space on which you can click. It will play the sentence with a specific expressiveness depending on its location.
    \item Familiarize yourself with it and listen to different sentences with a different expressiveness.
    \item Then for the experiment, use headphones to hear well, and be in a quiet environment where you will not be bothered.
    \item You will be asked to listen to a reference audio sample and find the red point in the 2D space that you feel to be the closest in expressiveness.
    \item Be aware that expressiveness varies continuously in the entire 2D space.
    \item You can click as much as you like on the 2D space and replay a sample. When you are satisfied with your choice, click on submit.
    \item There are two different versions, in the first one, the sentence is the same in the reference and in the 2D space. In the second, they are not. You just need to select the red point that in your opinion has the closest expressiveness.
    \item It would be great if you could do this for a set of 15 samples in each level. You can see your evolution on the page.

\end{itemize}

A number of 25 and 26 people participated in variants 1 and 2 of the experiment, respectively. We collected a total of 488 and 326 answers.\\

\begin{figure}[h]
  \centering
  \includegraphics[width=1\linewidth]{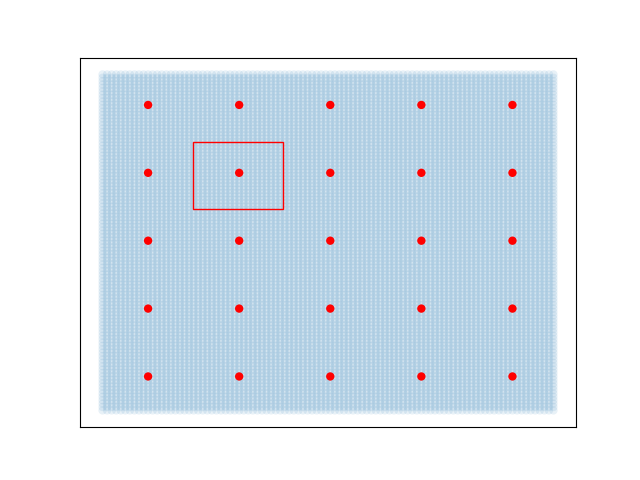}
  \caption{2D space fractionned in a 5x5 grid for the perceptual experiment. The red points are the possible positions of the reference in the space, the red rectangle is the selected case.}
\label{fig:grid}
\end{figure}

\subsection{Evaluation}

\paragraph{Controllability score}
To quantify how well the participants are able to produce a desired expressiveness, we compute an average euclidean distance between the selected point  and its true location.

Inspired by the omnipresent 5-point scales in the field of perceptual assessment, such as MOS tests, we choose to discretize the 2D space in a five-by-five grid, as shown in Figure~\ref{fig:grid}.
Indeed a continuous scale could be overwhelming for participants and let them unsure about their decision. The unit of distance is that between a red point and its neighbour along the horizontal axis.

We use a random baseline to assess the level of a non-controllability of the system in terms of expressiveness. In other words, if a participant is not able to distinguish the differences in expressiveness of different samples, we assume that he would not be able to select the correct location of the expressiveness of the reference, and would answer randomly.

\paragraph{Results and Discussion}

\begin{figure}[h]
  \centering
  \includegraphics[width=\linewidth]{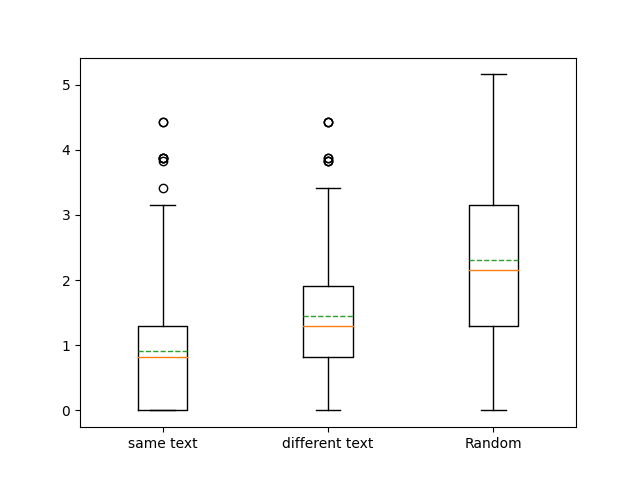}
  \caption{Boxplots of the distances between participant choices and true location of the reference (lower is better). From left to right: 1) results of the first variant of the experiment for which the text synthesized is the same for the reference and the latent space, 2) results of the second variant for which the text synthesized is different for the reference and the latent space, 3) a random baseline}
\label{fig:boxplot_icetalk_methods}
\end{figure}

Figure~\ref{fig:boxplot_icetalk_methods} shows the distributions of the distances between participant answers and true location of references in the 2D space. The two variants (with same text and different text) are on the left and the random baseline is on the right. The average distances with 95\% confidence intervals of the three distributions are respectively: $0.908\pm0.083$, $1.448 \pm 0.103$ and $2.314 \pm 0.007$.

\begin{figure}[h]
  \includegraphics[width=\linewidth]{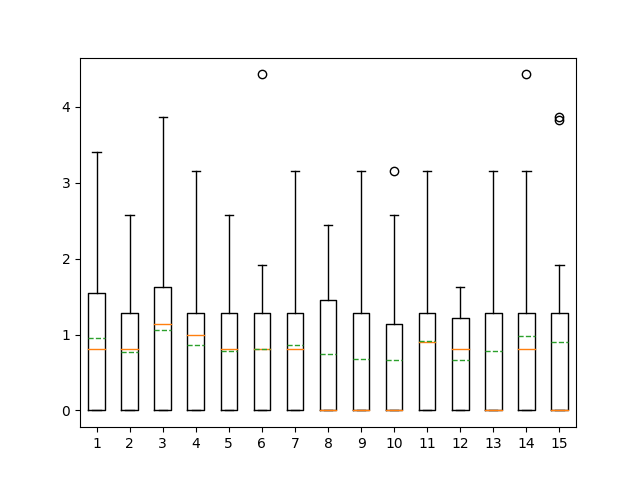}
   \includegraphics[width=\linewidth]{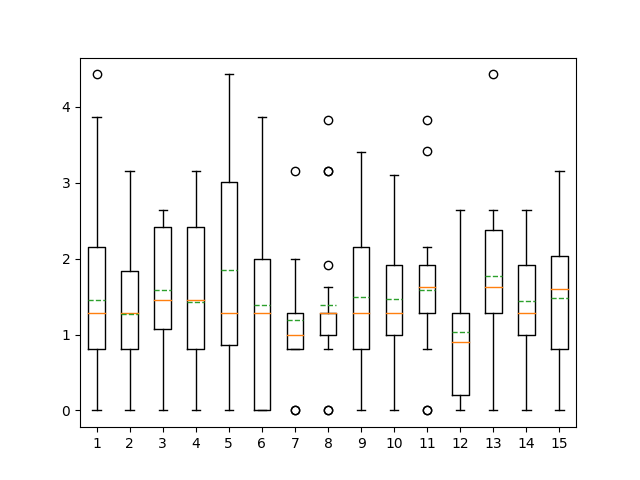}
  \caption{Boxplots of the distances between participant choices and true location of the reference by index until the 15th answer of participants  for variant 1 (top) and 2 (bottom) of the experiment.
  }
  \label{fig:boxplot_perceptual_exp_icetalk_by_idx}
\end{figure}


The second version was considered much more difficult by participants. 
For the first task, it is possible to listen to every detail of the intonation to detect if the sentence is the same.  That strategy is not possible for the second one in which only an abstract notion of expressiveness has to be imagined. 

Also, the speech rate is more difficult to compare between two different sentences than for the same sentence.
Generally speaking, when there is not the same number of syllables, it is more difficult to compare the melody and the rhythm of the sentences.

The cues mentioned by participants include intonation, tonic accent, speech rate and rhythm.

\begin{figure}[h]
  \includegraphics[width=\linewidth]{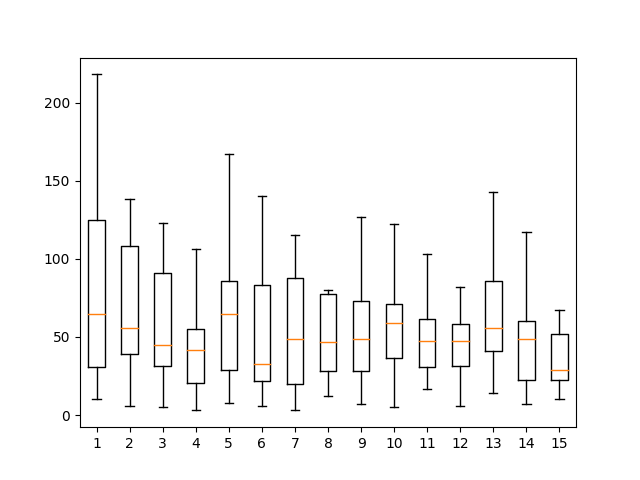}
   \includegraphics[width=\linewidth]{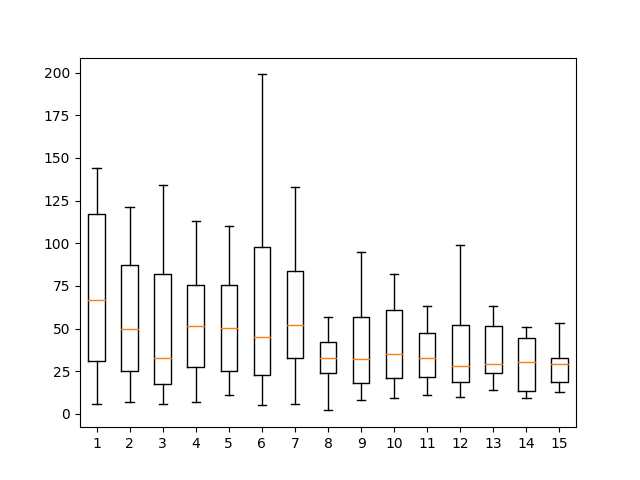}
  \caption{Boxplots of the durations for participant to answer by index until the 15th answer of participants  for variant 1 (top) and 2 (bottom) of the experiment.
  }
  \label{fig:boxplot_perceptual_exp_icetalk_by_idx_duration}
\end{figure}

We can see in Figure~\ref{fig:boxplot_perceptual_exp_icetalk_by_idx_duration} that, over time, participants are progressively more constant in the duration and with a lower median duration.
Outliers were discarded for plotting because they were too far from the distribution. The maximum is above 17500 seconds. We believe these outliers are due to a pauses taken by participants during the test. Also, the means are influenced by these outliers, and are therefore not plot in the figure.

A least square linear regression on the medians shows that it decreases with a slope of $-0.767$ s/task for the first variant and $-2.086$ s/task for the second.
The two-sided p-value for a hypothesis test whose null hypothesis is that the slope is zero are respectively $0.21$ and $0.0004$. We can therefore reject the null hypothesis in the second case but not in the first. 

Participants know more how to do it after several samples. They can guess where they have to search. They can establish a strategy as they understand how the space is structured. Therefore they feel like it is easier and they can make a choice faster because they hesitate less.

However, the evolution of average scores do not seem to improve or decline over time. A least square linear regression on the average scores show slopes close to zero for both variant 1 and 2 (respectively $-0.005$ and $0.0001$ s/task). 
The two-sided p-value for a hypothesis test whose null hypothesis is that the slope is zero are respectively $0.496$ and $0.930$. It indicates strong evidence that the slope is zero, i.e., the evolution of average scores remains stable.



\section{Summary and Conclusions}
\label{sec:ccl}

This paper presented a methodology for automatically building latent spaces related to expressiveness in speech data, for the purpose of controlling expressiveness in speech synthesis without referring to expert-based models of expressiveness. We then studied the relationships between such latent spaces and known audio features, to obtain a sense of the impact of such audio features on the styles expressed. This analysis consisted in an approximation of audio features from embeddings by linear regression. The accuracy of approximations was then evaluated in terms of correlations with ground truth.

The gradient of these linear approximations were computed to extract the information from variations of audio features in speech. By visualizing these gradients along with the embeddings, we observed the trends of audio features in latent spaces. 

A perceptual experiment was designed to evaluate the controllability of an Expressive TTS model based on these latent spaces. For that purpose, a set of reference utterances were synthesized with expressive control taken from discrete points in the 2D-reduced latent space. Test utterances were also synthesized with expressive control taken from a 5-by-5 grid on this 2D space.  Participants were then asked to search this 2D grid for the test utterance corresponding to the expressiveness of a reference utterance. An average distance on the grid was computed and compared to a random baseline. Two variants of the task were presented to participants: in the first one, the same sentence was used for the reference and test utterances, while in the second they were different. Results show that the average distance is lower for the first task than for the second, and that they are both lower than the random baseline.

\section{Perspectives}

We presented a 2D interface in which we can explore a space of expressiveness. It could be interesting to investigate ways to control more vocal characteristics, and independently when it is consistent and possible.
Several types of controls could be investigated depending on the nature of the variables. For some variables, the control could consist of a set of choices, e.g., male/female, or a list of speaker identities.

We also could imagine to have two separate 2D spaces. One would be dedicated to a speaker identity, i.e., a space organizing voice timbers. And the second would, e.g., the 2D space of expressiveness presented in this paper.
This kind of application needs frameworks able to \textit{disentangle} speech characteristics and factorize information corresponding to different phenomena, such as phonetics, speaker characteristics and expressiveness in the generated speech.

In the idea of having more and more general systems, the research results of this paper that focus on English language could be adapted to obtain a system able to work with several languages. This could be considered as one more aspect of speech that needs to be \textit{factorized} with others mentioned in previous paragraph.

There are also possibilities of controlling the evolution of speech characteristics inside a sentence, referred to as \textit{fine-grained} control that could be interesting to investigate. Currently, this aspect is mostly present in \textit{prosody transfer} task and is not subject to a control involving a human choosing what intonation, tonic accent or voice quality he would like to hear at different parts of a sentence. The difficulty would be to select the relevant characteristics that a sound designer would want to control and design an intuitive interface to control them. 

The different possibilities in this area would be interesting for, e.g., video games producer for the development of virtual characters with expressive voices, for animation movies, synthetic audiobooks, or in the advertisement sector.





\begin{acks}
No\'e Tits  is funded through a FRIA grant (Fonds pour la Formation \`a la Recherche dans l'Industrie et l'Agriculture, Belgium)\\
\url{https://app.dimensions.ai/details/grant/grant.8951680}
\end{acks}

\bibliographystyle{ACM-Reference-Format}
\bibliography{biblio}


\begin{thebibliography}{23}


\ifx \showCODEN    \undefined \def \showCODEN     #1{\unskip}     \fi
\ifx \showDOI      \undefined \def \showDOI       #1{#1}\fi
\ifx \showISBNx    \undefined \def \showISBNx     #1{\unskip}     \fi
\ifx \showISBNxiii \undefined \def \showISBNxiii  #1{\unskip}     \fi
\ifx \showISSN     \undefined \def \showISSN      #1{\unskip}     \fi
\ifx \showLCCN     \undefined \def \showLCCN      #1{\unskip}     \fi
\ifx \shownote     \undefined \def \shownote      #1{#1}          \fi
\ifx \showarticletitle \undefined \def \showarticletitle #1{#1}   \fi
\ifx \showURL      \undefined \def \showURL       {\relax}        \fi
\providecommand\bibfield[2]{#2}
\providecommand\bibinfo[2]{#2}
\providecommand\natexlab[1]{#1}
\providecommand\showeprint[2][]{arXiv:#2}

\bibitem[\protect\citeauthoryear{Akuzawa, Iwasawa, and Matsuo}{Akuzawa
  et~al\mbox{.}}{2018}]%
        {expressive_speech_vae-18-akuzawa}
\bibfield{author}{\bibinfo{person}{Kei Akuzawa}, \bibinfo{person}{Yusuke
  Iwasawa}, {and} \bibinfo{person}{Yutaka Matsuo}.}
  \bibinfo{year}{2018}\natexlab{}.
\newblock \showarticletitle{Expressive Speech Synthesis via Modeling
  Expressions with Variational Autoencoder}. In \bibinfo{booktitle}{\emph{Proc.
  Interspeech 2018}}. \bibinfo{pages}{3067--3071}.
\newblock
\urldef\tempurl%
\url{https://doi.org/10.21437/Interspeech.2018-1113}
\showDOI{\tempurl}


\bibitem[\protect\citeauthoryear{Burkhardt and Campbell}{Burkhardt and
  Campbell}{2014}]%
        {emotional_speech_synthesis-14-Burkhardt}
\bibfield{author}{\bibinfo{person}{Felix Burkhardt} {and} \bibinfo{person}{Nick
  Campbell}.} \bibinfo{year}{2014}\natexlab{}.
\newblock \showarticletitle{Emotional speech synthesis}.
\newblock In \bibinfo{booktitle}{\emph{The Oxford Handbook of Affective
  Computing}}. \bibinfo{publisher}{Oxford University Press New York},
  \bibinfo{pages}{286}.
\newblock


\bibitem[\protect\citeauthoryear{Eyben, Scherer, Schuller, Sundberg, Andr\'e,
  Busso, Devillers, Epps, Laukka, Narayanan, et~al\mbox{.}}{Eyben
  et~al\mbox{.}}{2016}]%
        {egemaps-16-eyben}
\bibfield{author}{\bibinfo{person}{Florian Eyben}, \bibinfo{person}{Klaus~R
  Scherer}, \bibinfo{person}{Bj{\"o}rn~W Schuller}, \bibinfo{person}{Johan
  Sundberg}, \bibinfo{person}{Elisabeth Andr\'e}, \bibinfo{person}{Carlos
  Busso}, \bibinfo{person}{Laurence~Y Devillers}, \bibinfo{person}{Julien
  Epps}, \bibinfo{person}{Petri Laukka}, \bibinfo{person}{Shrikanth~S
  Narayanan}, {et~al\mbox{.}}} \bibinfo{year}{2016}\natexlab{}.
\newblock \showarticletitle{{The Geneva minimalistic acoustic parameter set
  (GeMAPS) for voice research and affective computing}}.
\newblock \bibinfo{journal}{\emph{IEEE Transactions on Affective Computing}}
  \bibinfo{volume}{7}, \bibinfo{number}{2} (\bibinfo{year}{2016}),
  \bibinfo{pages}{190--202}.
\newblock


\bibitem[\protect\citeauthoryear{Henter, Lorenzo-Trueba, Wang, and
  Yamagishi}{Henter et~al\mbox{.}}{2018}]%
        {unsupervised_controllable_speech-18-henter}
\bibfield{author}{\bibinfo{person}{Gustav~Eje Henter}, \bibinfo{person}{Jaime
  Lorenzo-Trueba}, \bibinfo{person}{Xin Wang}, {and} \bibinfo{person}{Junichi
  Yamagishi}.} \bibinfo{year}{2018}\natexlab{}.
\newblock \showarticletitle{{Deep Encoder-Decoder Models for Unsupervised
  Learning of Controllable Speech Synthesis}}.
\newblock \bibinfo{journal}{\emph{arXiv preprint arXiv:1807.11470}}
  (\bibinfo{year}{2018}).
\newblock


\bibitem[\protect\citeauthoryear{Hsu, Zhang, Weiss, Zen, Wu, Wang, Cao, Jia,
  Chen, Shen, et~al\mbox{.}}{Hsu et~al\mbox{.}}{2018}]%
        {generative_controllable_speech-18-hsu}
\bibfield{author}{\bibinfo{person}{Wei-Ning Hsu}, \bibinfo{person}{Yu Zhang},
  \bibinfo{person}{Ron~J Weiss}, \bibinfo{person}{Heiga Zen},
  \bibinfo{person}{Yonghui Wu}, \bibinfo{person}{Yuxuan Wang},
  \bibinfo{person}{Yuan Cao}, \bibinfo{person}{Ye Jia},
  \bibinfo{person}{Zhifeng Chen}, \bibinfo{person}{Jonathan Shen},
  {et~al\mbox{.}}} \bibinfo{year}{2018}\natexlab{}.
\newblock \showarticletitle{{Hierarchical Generative Modeling for Controllable
  Speech Synthesis}}.
\newblock \bibinfo{journal}{\emph{arXiv preprint arXiv:1810.07217}}
  (\bibinfo{year}{2018}).
\newblock


\bibitem[\protect\citeauthoryear{Ito}{Ito}{2017}]%
        {ljspeech-17-keithito}
\bibfield{author}{\bibinfo{person}{Keith Ito}.}
  \bibinfo{year}{2017}\natexlab{}.
\newblock \bibinfo{title}{The LJ Speech Dataset}.
\newblock
  \bibinfo{howpublished}{\url{https://keithito.com/LJ-Speech-Dataset/}}.
\newblock


\bibitem[\protect\citeauthoryear{Karlapati, Moinet, Joly, Klimkov,
  Sáez-Trigueros, and Drugman}{Karlapati et~al\mbox{.}}{2020}]%
        {copycat-20-drugman}
\bibfield{author}{\bibinfo{person}{Sri Karlapati}, \bibinfo{person}{Alexis
  Moinet}, \bibinfo{person}{Arnaud Joly}, \bibinfo{person}{Viacheslav Klimkov},
  \bibinfo{person}{Daniel Sáez-Trigueros}, {and} \bibinfo{person}{Thomas
  Drugman}.} \bibinfo{year}{2020}\natexlab{}.
\newblock \showarticletitle{{CopyCat: Many-to-Many Fine-Grained Prosody
  Transfer for Neural Text-to-Speech}}. In \bibinfo{booktitle}{\emph{Proc.
  Interspeech 2020}}. \bibinfo{pages}{4387--4391}.
\newblock
\urldef\tempurl%
\url{https://doi.org/10.21437/Interspeech.2020-1251}
\showDOI{\tempurl}


\bibitem[\protect\citeauthoryear{Klimkov, Ronanki, Rohnke, and Drugman}{Klimkov
  et~al\mbox{.}}{2019}]%
        {fine_grained_prosody_transfer-19-drugman}
\bibfield{author}{\bibinfo{person}{Viacheslav Klimkov},
  \bibinfo{person}{Srikanth Ronanki}, \bibinfo{person}{Jonas Rohnke}, {and}
  \bibinfo{person}{Thomas Drugman}.} \bibinfo{year}{2019}\natexlab{}.
\newblock \showarticletitle{{Fine-Grained Robust Prosody Transfer for
  Single-Speaker Neural Text-To-Speech}}. In \bibinfo{booktitle}{\emph{Proc.
  Interspeech 2019}}. \bibinfo{pages}{4440--4444}.
\newblock
\urldef\tempurl%
\url{https://doi.org/10.21437/Interspeech.2019-2571}
\showDOI{\tempurl}


\bibitem[\protect\citeauthoryear{{Kubichek}}{{Kubichek}}{1993}]%
        {MCD}
\bibfield{author}{\bibinfo{person}{R. {Kubichek}}.}
  \bibinfo{year}{1993}\natexlab{}.
\newblock \showarticletitle{Mel-cepstral distance measure for objective speech
  quality assessment}. In \bibinfo{booktitle}{\emph{Proceedings of IEEE Pacific
  Rim Conference on Communications Computers and Signal Processing}},
  Vol.~\bibinfo{volume}{1}. \bibinfo{pages}{125--128 vol.1}.
\newblock


\bibitem[\protect\citeauthoryear{Nakatani, Amano, Irino, Ishizuka, and
  Kondo}{Nakatani et~al\mbox{.}}{2008}]%
        {VDE}
\bibfield{author}{\bibinfo{person}{Tomohiro Nakatani},
  \bibinfo{person}{Shigeaki Amano}, \bibinfo{person}{Toshio Irino},
  \bibinfo{person}{Kentaro Ishizuka}, {and} \bibinfo{person}{Tadahisa Kondo}.}
  \bibinfo{year}{2008}\natexlab{}.
\newblock \showarticletitle{A method for fundamental frequency estimation and
  voicing decision: Application to infant utterances recorded in real
  acoustical environments}.
\newblock \bibinfo{journal}{\emph{Speech Communication}} \bibinfo{volume}{50},
  \bibinfo{number}{3} (\bibinfo{year}{2008}), \bibinfo{pages}{203--214}.
\newblock


\bibitem[\protect\citeauthoryear{Raitio, Rasipuram, and Castellani}{Raitio
  et~al\mbox{.}}{2020}]%
        {controllable_NTTS-20-raitio}
\bibfield{author}{\bibinfo{person}{Tuomo Raitio}, \bibinfo{person}{Ramya
  Rasipuram}, {and} \bibinfo{person}{Dan Castellani}.}
  \bibinfo{year}{2020}\natexlab{}.
\newblock \showarticletitle{Controllable neural text-to-speech synthesis using
  intuitive prosodic features}.
\newblock \bibinfo{journal}{\emph{arXiv preprint arXiv:2009.06775}}
  (\bibinfo{year}{2020}).
\newblock


\bibitem[\protect\citeauthoryear{Shechtman and Sorin}{Shechtman and
  Sorin}{2019}]%
        {seq2seq_prosody_modif-19-shechtman}
\bibfield{author}{\bibinfo{person}{Slava Shechtman} {and} \bibinfo{person}{Alex
  Sorin}.} \bibinfo{year}{2019}\natexlab{}.
\newblock \showarticletitle{Sequence to sequence neural speech synthesis with
  prosody modification capabilities}.
\newblock \bibinfo{journal}{\emph{arXiv preprint arXiv:1909.10302}}
  (\bibinfo{year}{2019}).
\newblock


\bibitem[\protect\citeauthoryear{Skerry-Ryan, Battenberg, Xiao, Wang, Stanton,
  Shor, Weiss, Clark, and Saurous}{Skerry-Ryan et~al\mbox{.}}{2018}]%
        {tacotron_prosody-18-skerry}
\bibfield{author}{\bibinfo{person}{RJ Skerry-Ryan}, \bibinfo{person}{Eric
  Battenberg}, \bibinfo{person}{Ying Xiao}, \bibinfo{person}{Yuxuan Wang},
  \bibinfo{person}{Daisy Stanton}, \bibinfo{person}{Joel Shor},
  \bibinfo{person}{Ron Weiss}, \bibinfo{person}{Rob Clark}, {and}
  \bibinfo{person}{Rif~A Saurous}.} \bibinfo{year}{2018}\natexlab{}.
\newblock \showarticletitle{Towards End-to-End Prosody Transfer for Expressive
  Speech Synthesis with Tacotron}. In \bibinfo{booktitle}{\emph{International
  Conference on Machine Learning}}. \bibinfo{pages}{4693--4702}.
\newblock


\bibitem[\protect\citeauthoryear{Tachibana, Uenoyama, and Aihara}{Tachibana
  et~al\mbox{.}}{2018}]%
        {dctts-17-tachibana}
\bibfield{author}{\bibinfo{person}{Hideyuki Tachibana},
  \bibinfo{person}{Katsuya Uenoyama}, {and} \bibinfo{person}{Shunsuke Aihara}.}
  \bibinfo{year}{2018}\natexlab{}.
\newblock \showarticletitle{Efficiently trainable text-to-speech system based
  on deep convolutional networks with guided attention}. In
  \bibinfo{booktitle}{\emph{2018 IEEE International Conference on Acoustics,
  Speech and Signal Processing (ICASSP)}}. IEEE, \bibinfo{pages}{4784--4788}.
\newblock


\bibitem[\protect\citeauthoryear{Taigman, Wolf, Polyak, and Nachmani}{Taigman
  et~al\mbox{.}}{2017}]%
        {voiceloop-17-taigman}
\bibfield{author}{\bibinfo{person}{Yaniv Taigman}, \bibinfo{person}{Lior Wolf},
  \bibinfo{person}{Adam Polyak}, {and} \bibinfo{person}{Eliya Nachmani}.}
  \bibinfo{year}{2017}\natexlab{}.
\newblock \showarticletitle{Voiceloop: Voice fitting and synthesis via a
  phonological loop}.
\newblock \bibinfo{journal}{\emph{arXiv preprint arXiv:1707.06588}}
  (\bibinfo{year}{2017}).
\newblock


\bibitem[\protect\citeauthoryear{Tits, Haddad, and Dutoit}{Tits
  et~al\mbox{.}}{2021}]%
        {ICE_TALK2-21-tits}
\bibfield{author}{\bibinfo{person}{Noé Tits}, \bibinfo{person}{Kevin~El
  Haddad}, {and} \bibinfo{person}{Thierry Dutoit}.}
  \bibinfo{year}{2021}\natexlab{}.
\newblock \showarticletitle{ICE-Talk 2: Interface for Controllable Expressive
  TTS with perceptual assessment tool}.
\newblock \bibinfo{journal}{\emph{Software Impacts}} (\bibinfo{year}{2021}),
  \bibinfo{pages}{100055}.
\newblock
\showISSN{2665-9638}
\urldef\tempurl%
\url{https://doi.org/10.1016/j.simpa.2021.100055}
\showDOI{\tempurl}


\bibitem[\protect\citeauthoryear{Tits, Wang, Haddad, Pagel, and Dutoit}{Tits
  et~al\mbox{.}}{2019}]%
        {visualization-19-tits}
\bibfield{author}{\bibinfo{person}{No\'e Tits}, \bibinfo{person}{Fengna Wang},
  \bibinfo{person}{Kevin~El Haddad}, \bibinfo{person}{Vincent Pagel}, {and}
  \bibinfo{person}{Thierry Dutoit}.} \bibinfo{year}{2019}\natexlab{}.
\newblock \showarticletitle{{Visualization and Interpretation of Latent Spaces
  for Controlling Expressive Speech Synthesis through Audio Analysis}}. In
  \bibinfo{booktitle}{\emph{Proc. Interspeech 2019}}.
  \bibinfo{pages}{4475--4479}.
\newblock
\urldef\tempurl%
\url{https://doi.org/10.21437/Interspeech.2019-1426}
\showDOI{\tempurl}


\bibitem[\protect\citeauthoryear{van~den Oord, Dieleman, Zen, Simonyan,
  Vinyals, Graves, Kalchbrenner, Senior, and Kavukcuoglu}{van~den Oord
  et~al\mbox{.}}{2016}]%
        {wavenet-16-vandenoord}
\bibfield{author}{\bibinfo{person}{A{\"a}ron van~den Oord},
  \bibinfo{person}{Sander Dieleman}, \bibinfo{person}{Heiga Zen},
  \bibinfo{person}{Karen Simonyan}, \bibinfo{person}{Oriol Vinyals},
  \bibinfo{person}{Alex Graves}, \bibinfo{person}{Nal Kalchbrenner},
  \bibinfo{person}{Andrew~W. Senior}, {and} \bibinfo{person}{Koray
  Kavukcuoglu}.} \bibinfo{year}{2016}\natexlab{}.
\newblock \showarticletitle{{WaveNet: A Generative Model for Raw Audio}}. In
  \bibinfo{booktitle}{\emph{SSW}}.
\newblock


\bibitem[\protect\citeauthoryear{Wang, Skerry-Ryan, Stanton, Wu, Weiss, Jaitly,
  Yang, Xiao, Chen, Bengio, Le, Agiomyrgiannakis, Clark, and Saurous}{Wang
  et~al\mbox{.}}{2017}]%
        {tacotron-17-wang}
\bibfield{author}{\bibinfo{person}{Yuxuan Wang}, \bibinfo{person}{R.~J.
  Skerry-Ryan}, \bibinfo{person}{Daisy Stanton}, \bibinfo{person}{Yonghui Wu},
  \bibinfo{person}{Ron~J. Weiss}, \bibinfo{person}{Navdeep Jaitly},
  \bibinfo{person}{Zongheng Yang}, \bibinfo{person}{Ying Xiao},
  \bibinfo{person}{Zhifeng Chen}, \bibinfo{person}{Samy Bengio},
  \bibinfo{person}{Quoc~V. Le}, \bibinfo{person}{Yannis Agiomyrgiannakis},
  \bibinfo{person}{Rob Clark}, {and} \bibinfo{person}{Rif~A. Saurous}.}
  \bibinfo{year}{2017}\natexlab{}.
\newblock \showarticletitle{{Tacotron: Towards End-to-End Speech Synthesis}}.
  In \bibinfo{booktitle}{\emph{INTERSPEECH}}.
\newblock


\bibitem[\protect\citeauthoryear{Wang, Stanton, Zhang, Ryan, Battenberg, Shor,
  Xiao, Jia, Ren, and Saurous}{Wang et~al\mbox{.}}{2018}]%
        {tacotron_style_tokens-18-wang}
\bibfield{author}{\bibinfo{person}{Yuxuan Wang}, \bibinfo{person}{Daisy
  Stanton}, \bibinfo{person}{Yu Zhang}, \bibinfo{person}{RJ-Skerry Ryan},
  \bibinfo{person}{Eric Battenberg}, \bibinfo{person}{Joel Shor},
  \bibinfo{person}{Ying Xiao}, \bibinfo{person}{Ye Jia}, \bibinfo{person}{Fei
  Ren}, {and} \bibinfo{person}{Rif~A Saurous}.}
  \bibinfo{year}{2018}\natexlab{}.
\newblock \showarticletitle{Style Tokens: Unsupervised Style Modeling, Control
  and Transfer in End-to-End Speech Synthesis}. In
  \bibinfo{booktitle}{\emph{International Conference on Machine Learning}}.
  \bibinfo{pages}{5180--5189}.
\newblock


\bibitem[\protect\citeauthoryear{Watts, Henter, Merritt, Wu, and King}{Watts
  et~al\mbox{.}}{2016}]%
        {hmms_to_dnns-16-watts}
\bibfield{author}{\bibinfo{person}{Oliver Watts}, \bibinfo{person}{Gustav~Eje
  Henter}, \bibinfo{person}{Thomas Merritt}, \bibinfo{person}{Zhizheng Wu},
  {and} \bibinfo{person}{Simon King}.} \bibinfo{year}{2016}\natexlab{}.
\newblock \showarticletitle{{From HMMs to DNNs: where do the improvements come
  from?}}. In \bibinfo{booktitle}{\emph{Acoustics, Speech and Signal Processing
  (ICASSP), 2016 IEEE International Conference on}}. IEEE,
  \bibinfo{pages}{5505--5509}.
\newblock


\bibitem[\protect\citeauthoryear{Zen, Senior, and Schuster}{Zen
  et~al\mbox{.}}{2013}]%
        {stat_param_speech_synthesis_dnn-13-zen}
\bibfield{author}{\bibinfo{person}{Heiga Zen}, \bibinfo{person}{Andrew Senior},
  {and} \bibinfo{person}{Mike Schuster}.} \bibinfo{year}{2013}\natexlab{}.
\newblock \showarticletitle{Statistical parametric speech synthesis using deep
  neural networks}. In \bibinfo{booktitle}{\emph{Acoustics, Speech and Signal
  Processing (ICASSP), 2013 IEEE International Conference on}}. IEEE,
  \bibinfo{pages}{7962--7966}.
\newblock


\bibitem[\protect\citeauthoryear{Zen, Tokuda, and Black}{Zen
  et~al\mbox{.}}{2009}]%
        {statistical_param_speech_syn-09-zen}
\bibfield{author}{\bibinfo{person}{Heiga Zen}, \bibinfo{person}{Keiichi
  Tokuda}, {and} \bibinfo{person}{Alan~W Black}.}
  \bibinfo{year}{2009}\natexlab{}.
\newblock \showarticletitle{Statistical parametric speech synthesis}.
\newblock \bibinfo{journal}{\emph{Speech Communication}} \bibinfo{volume}{51},
  \bibinfo{number}{11} (\bibinfo{year}{2009}), \bibinfo{pages}{1039--1064}.
\newblock


\end{thebibliography}

\appendix




\end{document}